\documentclass[journal]{IEEEtran}
\usepackage{amssymb}
\ifCLASSINFOpdf
   \usepackage[pdftex]{graphicx}
\else
   \usepackage[dvips]{graphicx}
\fi
\usepackage{subfigure} 
\usepackage{float}
\usepackage{setspace}
\usepackage[cmex10]{amsmath}
\usepackage{mdwmath}
\usepackage{setspace}
\usepackage{multirow}
\usepackage{array}
\usepackage{cases,multicol}
\usepackage{epstopdf}
\usepackage[lined,boxed,commentsnumbered, ruled]{algorithm2e} 
\usepackage{amsthm} 

\usepackage{chemarrow}
\usepackage{extarrows}
\usepackage{color}
\allowdisplaybreaks[4]
\usepackage[colorlinks,
            linkcolor=black,
            anchorcolor=black,
            urlcolor=black,
            citecolor=black]{hyperref}
\usepackage[hyphenbreaks]{breakurl}
\usepackage{makecell}

\newcommand{\gainame}{{GAI}}
\newcommand{\ainame}{AI}
\newcommand{\qoename}{QoE}
\newcommand{\alertname}{HAS}

\newcommand{\systemname}{LLM-HAS}
\newcommand{\gdmname}{GDMs}

\newcommand{\drlname}{DRL}

\newcommand{\maname}{MA}
\newcommand{\faname}{FA}

\newcommand{\gdprname}{GDPR}

\newcommand{\domainname}{IoT-based HAS}

\newcommand{\iotname}{IoT}
\newcommand{\llmname}{LLM}
\newcommand{\llmsname}{LLMs}
\newcommand{\ddpgname}{DDPG}
\newcommand{\kpiname}{KPI}
\newcommand{\alertsname}{HASs}
\newcommand{\rlname}{RL}
\newcommand{\moename}{MoE}
\newcommand{\llmmoename}{LLM-enabled MoE}

\begin{document}
\title{Guiding IoT-Based Healthcare Alert Systems with Large Language Models}

\author{Yulan Gao,\IEEEmembership{~Member, IEEE,}
Ziqiang Ye,
Ming Xiao,\IEEEmembership{~Senior Member, IEEE,}
Yue Xiao,\IEEEmembership{~Member, IEEE,}\\
and Dong In Kim,\IEEEmembership{~Fellow, IEEE}
\thanks{Y. Gao and M. Xiao are with the Division of Information Science and Engineering, KTH Royal Institute of Technology, 100 44 Stockholm, Sweden  (e-mail: ylgao.math@gmail.com, mingx@kth.se).}
\thanks{Z. Ye and Y. Xiao are with the Key National Laboratory of Wireless Communications, University of Electronic Science and Technology of China (UESTC), Chengdu 611731, China (e-mail: yysxiaoyu@hotmail.com, xiaoyue@uestc.edu.cn).}
\thanks{D. I. Kim is with the Department of Electrical
and Computer Engineering, Sungkyunkwan University, Suwon 16419, South
Korea (e-mail: dongin@skku.edu).}
} 
\markboth{~}
{Shell \MakeLowercase{\textit{et al.}}: }
\maketitle

\begin{abstract}
Healthcare alert systems (\alertname{}) are undergoing rapid evolution, propelled by advancements in artificial intelligence (\ainame{}),  Internet of Things (IoT) technologies, and increasing health consciousness. 
Despite significant progress, a fundamental challenge remains: balancing the accuracy of personalized health alerts with stringent privacy protection in \alertname{} environments constrained by resources. 
To address this issue, we introduce a uniform framework, \systemname{}, which incorporates Large Language Models (\llmsname{}) into \alertname{} to significantly boost the accuracy, ensure user privacy, and enhance personalized health service, while also improving the subjective quality of experience (\qoename{}) for users. 
Our innovative framework leverages a Mixture of Experts (\moename{}) approach, augmented with \llmsname{}, to analyze users' personalized preferences and potential health risks from additional textual job descriptions.  
This analysis guides the selection of specialized Deep Reinforcement Learning (\drlname{}) experts, tasked with making precise health alerts. 
Moreover, \systemname{} can process Conversational User Feedback, which not only allows fine-tuning of \drlname{} but also deepen user engagement, thereby enhancing both the accuracy and personalization of health management strategies. 
Simulation results validate the effectiveness of the \systemname{} framework, highlighting its potential as a groundbreaking approach for employing generative \ainame{} (\gainame{}) to provide highly accurate and reliable alerts. 
\end{abstract}

\IEEEpeerreviewmaketitle

\section{Introduction}
\begin{figure*}[!t]
    \centering
    \includegraphics[width=\textwidth]{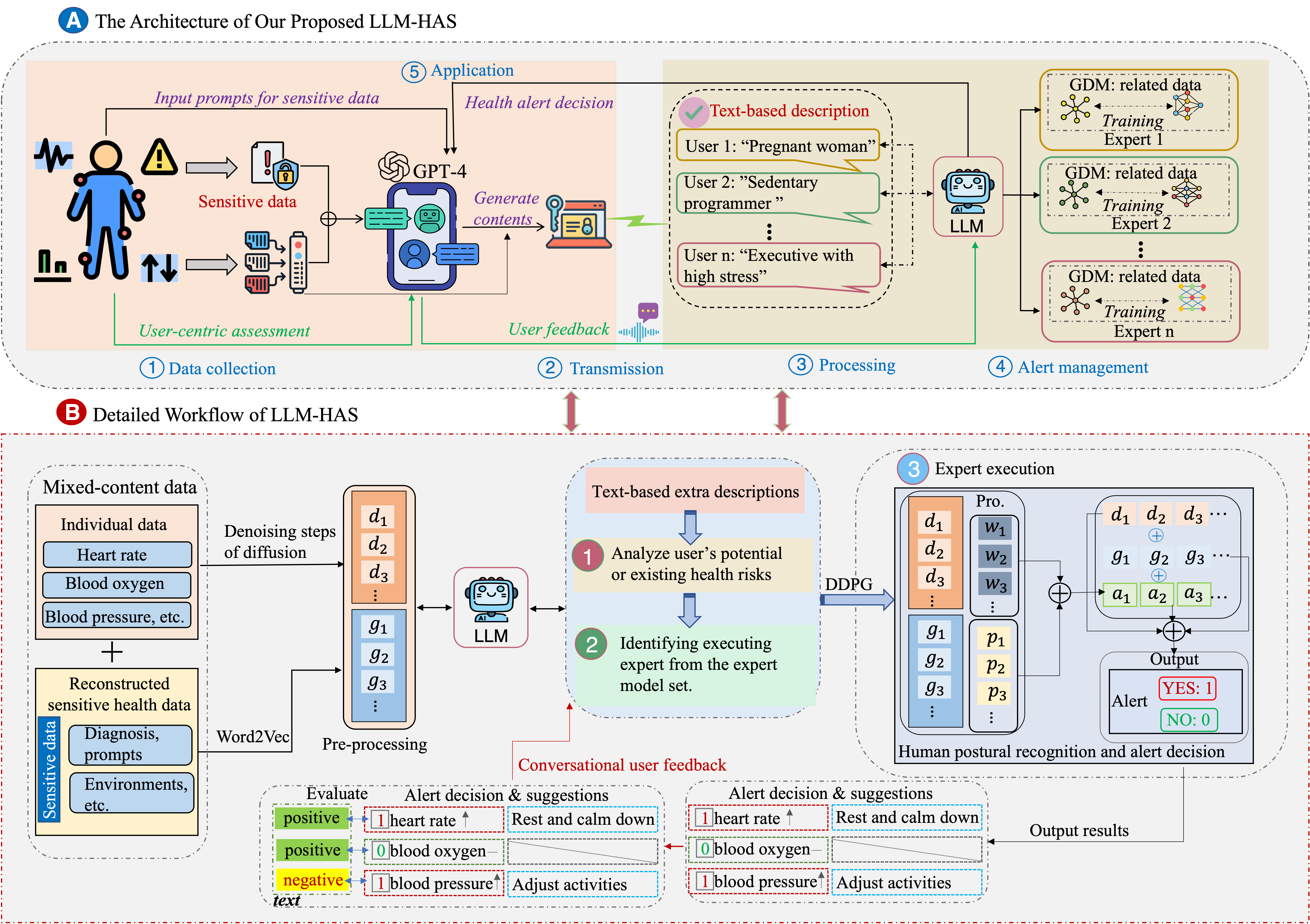}
    \caption{Architecture and  workflow of the proposed \systemname{} framework:  Once the system receives extra text-based descriptions of occupation or status, \llmname{} processes the data and selects the appropriate experts required to address the users' specific needs.}
    \label{fig:1}
\end{figure*}

The penetration of the Internet of Things (\iotname{}), big data, artificial intelligence (\ainame{}), cloud computing, and mobile applications into our society is revolutionizing healthcare. 
At the forefront of this revolution, \iotname{}-based Health Alert Systems (\domainname{}) showcase advanced capabilities in sensing, communication, computation, and control.
The architecture of a typical \domainname{} comprises several crucial modules: data collection, transmission, processing, alert management, and application, depicted in Fig. \ref{fig:1} (Part A). 
Together, these components collaborate to fulfill key performance indicators (KPI) such as real-time responsiveness, high accuracy, scalability, and user-friendliness, significantly advancing the capabilities beyond those of traditional \alertname{} applications \cite{esposito2018smart}.   

However, the effectiveness and reliability of these systems still heavily rely on the integrity and quality of the data collected, as well as on the implementation of sophisticated adaptive \ainame{} models. 
Amid escalating concerns about data security and personal privacy, stringent regulatory frameworks like the General Data Protection Regulation (\gdprname{}) by the European Union have been enacted to uphold robust data protection standards \cite{gdpr2018general}. 
Moreover, as the smart healthcare ecosystem continues to integrate a diverse array of sensors and multimodal data, stringent privacy regulations and the inherent complexity of medical data present formidable challenges. 
These restrictions limit access to crucial personal health data, thereby impeding the development of sophisticated algorithms essential for generating accurate healthcare alert decisions.

Furthermore, examining societal development through a health-focused lens reveals that different occupations expose people to unique health risks, a trend that might be seen as the {\em professionalization} of health issues \cite{hege2019impact}. 
For example, sedentary occupations such as programming and scientific research elevate risks for musculoskeletal and cardiovascular conditions, mental health struggles, eye strain, and nutritional deficiencies due to irregular eating. 
Thus, health management systems designed for programmers and researchers, for example, can monitor and send reminders for heart rate, blood pressure, sleep, and posture, enhancing the system’s precision, relevance, and efficacy. 
This targeted approach not only promotes healthier work habits but also mitigates the risks of chronic health conditions, proving indispensable for diverse professional groups. 
However, the integration of AI models, each tailored to specific tasks and datasets, into smart healthcare systems presents its own challenges. 
The rapid expansion of these technologies places significant demands on edge servers, straining their capacity due to the high requirements for training and deploying resources.

Fortunately, Generative \ainame{} (\gainame{}) presents a robust alternative or complement to traditional \ainame{} models within \alertname{}, adeptly addressing the  challenges previously outlined. 
\gainame{} models have garnered significant attention for their remarkable capacity to autonomously generate diverse content, including text, images, voice, and videos \cite{ooi2023potential}.
Large language models (\llmsname{}), exemplifies by OpenAI's GPT-4, launched in March 2023, represent a pinnacle in \gainame{} technology. 
These models serve as highly versatile tools, adept at machine translation, text generation, and semantic analysis, among other applications, and are specifically engineered to both comprehend and generate human language with remarkable efficacy \cite{kalyan2023survey}.  
The GPT-4 iOS client enhances accessibility with its voice-to-text feature, making it not only appealing to younger demographics but also exceptionally user--friendly for the elderly. 
Additionally, generative diffusion models (\gdmname{}) \cite{cao2024survey}, which consist of three key components--the forward process, the reverse process, and the sampling procedure--are designed to master a diffusion process that generates a probability distribution for a dataset, thereby facilitating the creation of new images. 

Building on above discussion,  we present the first \llmname{}-based \alertname{} (\systemname{}) framework, as depicted in Fig. \ref{fig:1}. 
Drawing inspiration from \cite{du2024mixture}, this system leverages \llmname{} (GPT-4) on the user side and incorporates a \llmname{}-enabled Mixture of Experts (\moename{}) framework on the edge server. 
In this framework, experts are specialized models, such as Deep Reinforcement Learning (\drlname{}) models, each trained to perform well on specific subsets of the user data. 
To facilitate the training of these \drlname{} models, we employ a \gdmname{} to generate synthetic data, thereby enriching the existing dataset and enhancing model performance. 
On the user side, the \systemname{} employs GPT-4 to reconstruct sensitive health data from user prompts, combining it with non-sensitive data to create a mixed-content dataset. 
This data is then denoised on the edge server using  \gdmname{}. 
The \llmmoename{} architecture processes this pre-processed data by first utilizing a \llmname{} to interpret the user's occupation and associated health risks from text-based descriptions{\footnote{Individuals can describe their occupation and current status in various ways, depending on how they perceive the privacy of their personal information. 
Descriptions for programmers could vary from \{I am programmer\} to \{I engaged in long working hours and sedentary mental work\}.}. 
It then selects the appropriate experts, specifically choosing the necessary \drlname{} model, to make personalized health alert decisions for the users. 
Additionally, to enhance the subjective quality of experience (QoE) for users, \systemname{} receives conversational user feedback on alert outcomes and fine-tunes the parameters of the \drlname{} algorithm accordingly.    

Overall, the novelty and contributions of this paper are summarized in the following aspects.
\begin{itemize}
\item We propose \systemname{}, the first framework that integrates \llmname{} with \alertname{} and incorporates \llmmoename{} approach, specifically designed to reduce the probability of missed alerts (\maname{}) and false alerts (\faname{}).   
To enhance privacy protection, we propose guidelines for uploading mixed-content data to the edge server, where the sensitive health data is reconstructed by GPT-4 from prompts. 
\item By implementing the \llmmoename{} framework on the edge server, we facilitate the collaboration of multiple \drlname{} models within the resource-constrained \alertname{}. 
These models, guided by an \llmname{}, work in concert to optimize health alert decisions. 
This approach not only accommodates the evolving statues of users but also to their diverse occupational profiles, ensuring tailored and dynamic responses.     
\item We introduce an \llmname{}-enabled conversational user feedback mechanism into the \systemname{}, designed to optimize the balance between user subjective \qoename{} and the accuracy of health alert decisions. 
It is important to note that this feedback mechanism leverages natural language to provide intuitive and subjective evaluation of alert outcomes. 
The detailed process is illustrated in Fig. \ref{fig:3}. 
Then, the edge server receives this feedback and dynamically fine-tunes parameters of \drlname{} algorithm.   
\item We verify the feasibility and effectiveness of the proposed \systemname{} framework through a practical case study using the Mobile Health (MHEALTH) dataset from the {\bf{UC Irvine Machine Learning Repository}}. 
Additionally, we present an in-depth analysis of the \systemname{} framework, elucidating its transformative potential for \domainname{}, empowered by \gainame{}. 
\end{itemize}

The rest of the paper is organized as follows. 
The overview of \domainname{} is presented in Section \ref{sec:overview}. 
Section \ref{sec:framework} outlines the proposed \systemname{} framework. 
Sections \ref{sec:simulations} and \ref{sec:future} outline the experiments of \systemname{} and future directions, respectively, after which this paper is concluded in Section \ref{sec:conclusion}. 

\section{Overview of \domainname{} }\label{sec:overview}
 IoT-based HASs are designed to monitor and analyze health data in real-time, utilizing advanced technologies to detect potential health issues and trigger timely alerts.
These systems operate on the principle of preventive healthcare, where immediate responses and early interventions are facilitated through continuous monitoring, ensuring enhanced patient safety and improved health management. 
In the realm of \domainname{}, the integration of \ainame{} has significantly enhanced operational capabilities and system evolution. 
Initially, these systems employed basic AI techniques for straightforward data analysis and alert generation, utilizing set thresholds to monitor essential health metrics like heart rates or glucose levels. 
Over time, the adoption of more advanced \ainame{} forms such as reinforcement learning (\rlname{}) allowed these systems to become not only reactive but also predictive, providing a more nuanced decision-making approach by analyzing previous alert outcomes to reduce inaccuracies \cite{naeem2021reinforcement}. 
This progression continued with the integration of \drlname{}, which leveraged deep neural networks to effectively analyze extensive data sets and learn from diverse inputs, including patient behavioral patterns, thus significantly enhancing predictive accuracy and operational efficiency \cite{wang2024deep}. 
The most recent advancement involves the integration of \llmsname{} in optimizing \alertsname{}, particularly in the realm of clinical decision support \cite{schwartz2024black,liu2024users}. 
Specifically, Ref. \cite{schwartz2024black} showcased how \llmsname{} effectively summarize feedback from healthcare providers on clinical alerts, improving the clarity and actionability of these alerts while reducing fatigue. 
Ref. \cite{liu2024users} detailed how \llmsname{} aid in infectious disease management by predicting outbreaks and individual patient risks, leading to faster and more accurate medical responses. 

Our paper builds on these foundations by integrating \llmsname{} into \domainname{}, extending their application beyond mere summarization and prediction to include dynamic interactions with users.
From an operational perspective, we adapt the \llmmoename{} framework on the edge server, enabling it to respond to varying user statuses and diverse occupational profiles effectively. 
Leveraging \llmsname{}' powerful reasoning capabilities, this paper introduces a novel framework that not only refines system algorithms but also enhances user experiences through conversational interactions. 
These engagements facilitate a deeper personalization of the system, significantly improving user subjective experience while simultaneously bolstering privacy and data integrity by processing sensitive data locally. 
This approach ensures that our \systemname{} is not only responsive but also respects the nuanced needs of each user.

\section{The Proposed \systemname{} Framework}\label{sec:framework}
This section presents the proposed \systemname{} framework, outlining key research challenges and detailing the implementation processes.   
\begin{figure*}[!t]
    \centering
    \includegraphics[width=\textwidth]{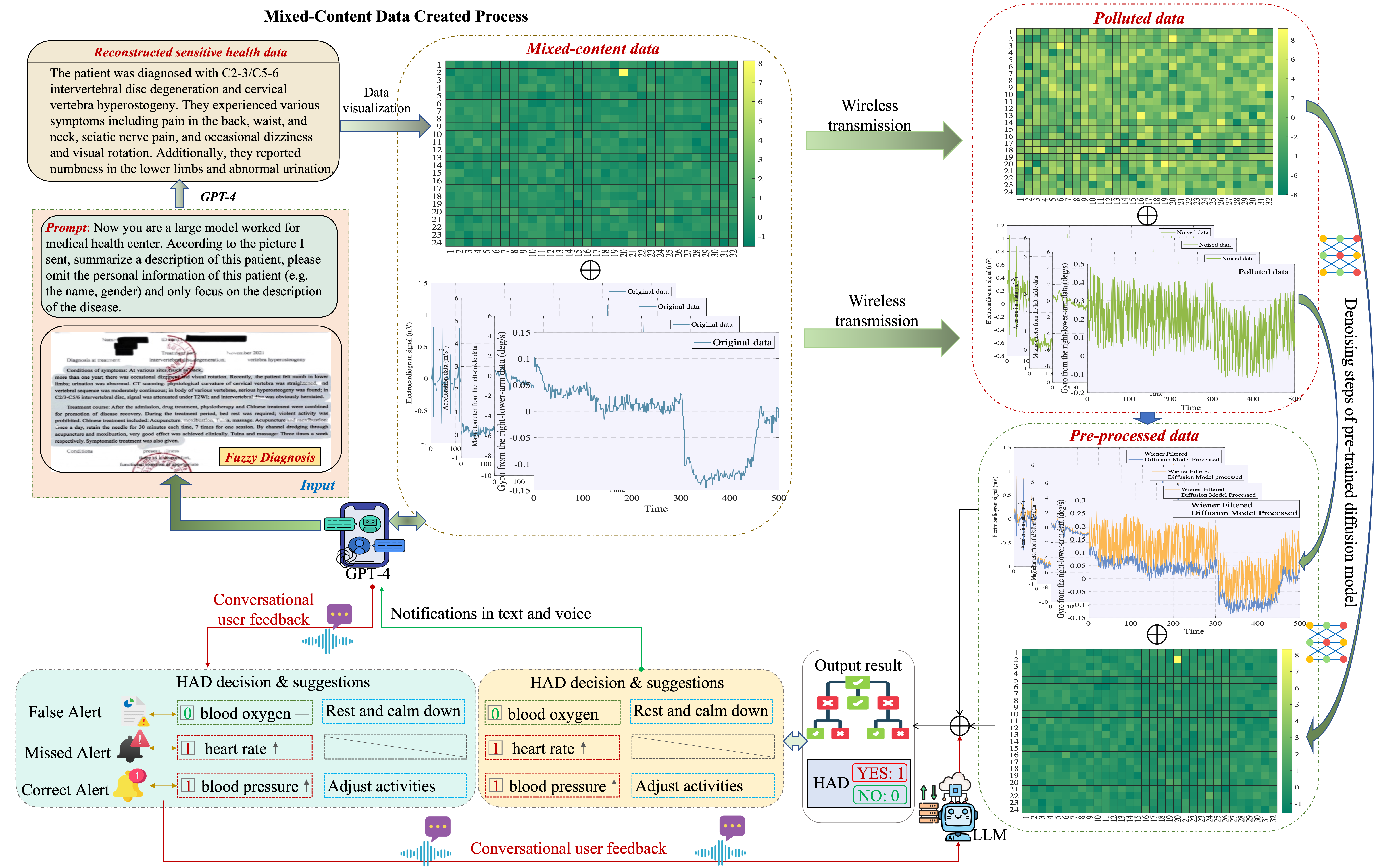}
    \caption{Processes for generating and pre-processing mixed-content data: a case study illustration.}
    \label{fig:2}
\end{figure*}

\subsection{Key Research Challenges}
{\bf\em 1) Privacy and data integrity.}
Privacy protection is crucial in smart healthcare, ensuring patient trust and legal compliance while enabling the safe use of advanced technologies.
However, the effectiveness and reliability of the \systemname{} framework hinge on the integrity and quality of the data collected. 
To ensure the system's success, it is essential to maintain the highest possible data quality and integrity alongside robust privacy protection. 
Thus, \systemname{} must handle sensitive health data responsible, ensuring that data privacy is maintained according to regulations like GDPR while integrating sensitive and non-sensitive data. 

{\bf\em 2) Adaptability.}
The \systemname{} framework must be highly adaptable, accommodating individual user profiles that encompass diverse medical histories, lifestyle factors, and evolving health conditions. 
To optimize adaptability, two critical strategies are employed. 
First, the system continuously fine-tunes its selected experts - namely, the \drlname{} models - using conversational user feedback. 
This process not only enhances the subjective \qoename{} for users but also improves the accuracy of health alert decisions. 
Second, the framework ensures its relevance and effectiveness by periodically updating the experts models to align with current health trends and user needs. 
The second strategy implementation based on the \llmmoename{} framework.

{\bf\em 3) Real-time processing.}
In the realm of smart healthcare, a timely response is crucial. The \systemname{} is designed to process data in real-time, enabling the delivery of immediate alerts. 
Given the advanced capabilities of \llmsname{} and \gdmname{}, resource intensity emerges as a significant concern, especially with the limited computing power and energy constraints of user devices. 
To tackle this, the \systemname{} framework delegates the resource-intensive tasks, including data denoising, human postural recognition, and health alert decision-making, to edge servers. 
As depicted in Fig. \ref{fig:1} (Part A), the \llmmoename{} framework,  deployed on these servers, is designed to optimize computational resources effectively. 
This approach minimizes delays and reduces energy consumption. 
The integration of \llmsname{} and \moename{} at the edge addresses the key challenges of real-time data processing and rapid decision-making, essential for delivering timely health alerts and interventions. 

{\bf\em 4) User engagement.}
User engagement poses a main challenge within the \systemname{} framework, necessitating a design that supports consistent interaction and responsiveness to user feedback.  
This continuous engagement is essential for the long-term effectiveness of \systemname{}, enhancing its ability to promote improved health outcomes. 
As illustrated in Fig. \ref{fig:1}, the framework utilizes an \llmname{}-enabled conversational user feedback mechanism to foster more intuitive interactions, thereby increasing accessibility for all users.
To ensure ongoing engagement, the system must provide timely, actionable health insights that encourage continuous interaction and build user trust. 
\begin{figure*}[!t]
    \centering
    \includegraphics[width=\textwidth]{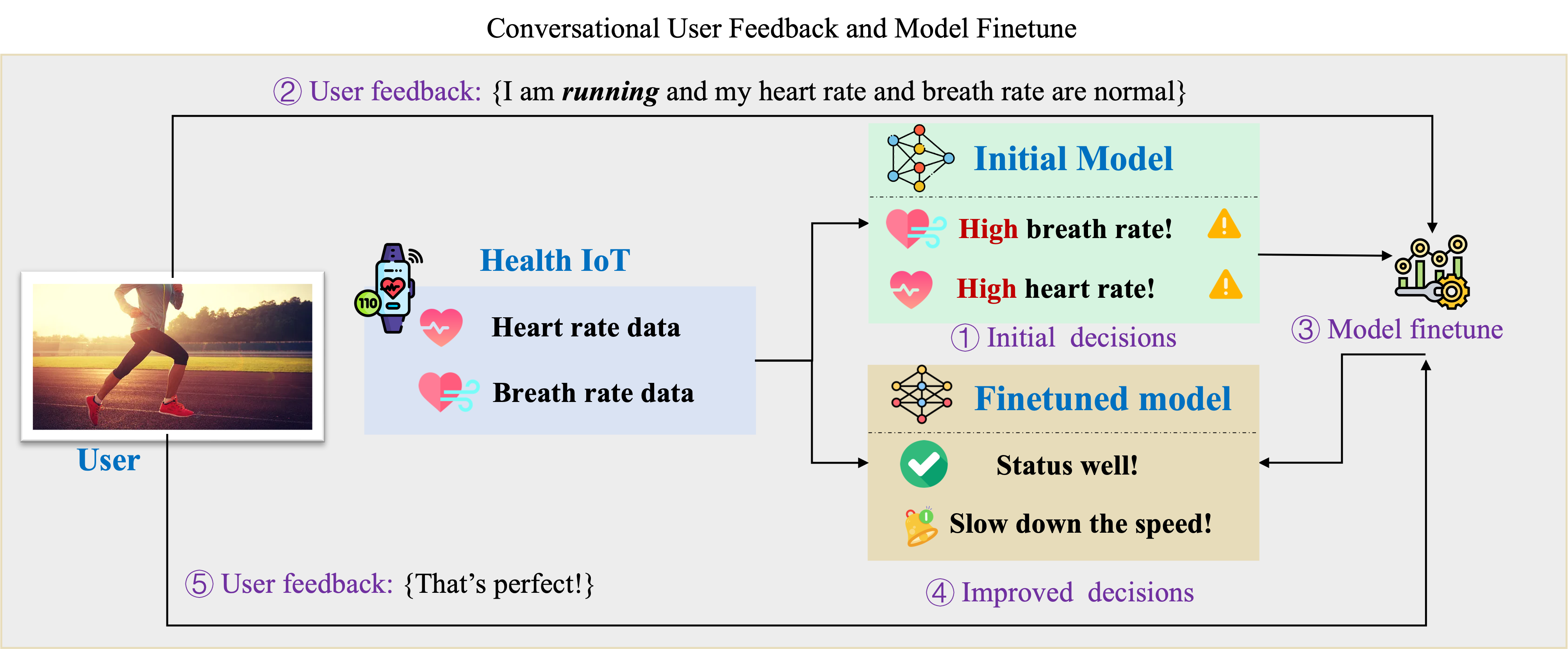}
    \caption{An example of dynamic conversational user feedback in the \systemname{} framework. }
    \label{fig:3}
\end{figure*}

\subsection{The Proposed \systemname{} Framework}
As illustrated in Fig. \ref{fig:1}, the proposed \systemname{} framework can be structured into four main modules. 
The first module handles mixed-content data acquisition on the user side, utilizing \llmname{}, such as GPT-4, to reconstruct the sensitive health data based on the raw data and user prompts.  
The second module employs the \llmmoename{} framework to analyze individual incoming data and efficiently identify the appropriate expert from the expert model set. 
This ensures that specialized knowledge is applied to each unique case. 
The third module, focused on health alert decision-making, employs the selected \ddpgname{} model to generate human postural-aware alerts.
This approach enhances the relevance and urgency of the health alerts provided.
The fourth module leverages \llmname{}-enabled conversational user feedback to dynamically fine-tune the \ddpgname{} model. 
By integrating multi-modal user inputs, including text and voice, this module not only refines the accuracy of the alerts but also significantly improves the user's subjective \qoename{}. 

\subsubsection{\llmname{}-Based Mixed-Content Data Creation}
To ensure the privacy protection and integrity of user health data, a robust approach involves using \llmname{} to reconstruct sensitive data, thereby preventing privacy leakage and potential GDPR violations associated with direct data transmission. 
Initially, wearable sensors gather human physiological parameters and transmit them to a hub node, typically a smartphone, tablet, or similar device. 
The hub node then discerns which data can be directly uploaded and identifies which data is sensitive. 
Subsequently, any sensitive data is locally extracted by the \llmname{} relying on the requirements of users,  ensuring enhanced privacy and data integrity. 
It is important to note that the requirements of users should avoid any sensitive prompts and be descriptive, aligning closely with the user's subjective preferences to ensure accuracy and confidentiality as much as possible.

As shown in Fig. \ref{fig:2}, the extraction and reconstruction of sensitive data labeled as {\em Diagnosis} by the \llmname{}-based mechanism is represented as $\mbox{LLM}\{{\boldsymbol p}\}\rightarrow {\boldsymbol r}$.
Here, $\mbox{LLM}\{\cdot\}$ denotes the inference process of the \llmname{}, ${\boldsymbol p}$ is the input requirement\footnote{For privacy protection, the diagnosis displayed in the figure is blurred; however, the actual input utilized consists of the clear and complete content of the diagnosis.} of user such as ``\{{\bf\em Now you are a large model worked for medical health center. According to the picture I sent, summarize a description of this patient, please omit the personal information of this patient (e.g. the name, gender) and only focus on the description of the disease.}\}'' 
The diagnosis reconstructed by GPT-4 is: ``\{{\bf\em The patient was diagnosed with C2-3/C5-6 intervertebral disc degeneration and cervical vertebra hyperostogeny. They experienced various symptoms including pain in the back, waist, and neck, sciatic nerve pain, and occasional dizziness and visual rotation. Additionally, they reported numbness in the lower limbs and abnormal urination.}\}'' This text, along with the physiological data collected by sensors, is then aggregated into a multi-modal mixed-content data.  This data is then transmitted to the edge server via wireless channels.  Moreover, all operations are executed locally on the hub node, thereby ensuring enhanced data security and robust privacy protection. 

\subsubsection{\llmmoename{} Framework}
In our \systemname{}, the initial step on the edge server involves refining the acquiring data using a pre-trained diffusion model. 
Following this initial purification, the text-particularly those components containing sensitive information-is vectorized utilizing the Word2Vec technique \cite{Text2vec}. 
Within the framework of \systemname{} (as depicted in Part B of Fig. \ref{fig:1}), the content generated by \llmname{} is effectively transformed into a vector representation, denoted as ${\boldsymbol g}=[g_1, g_2, \ldots, g_M]^T$. 
Concurrently, non-sensitive data is encapsulated separately and is represented by ${\boldsymbol d}=[d_1, d_2, \ldots, d_N]^T$. 
Here, $M$ and $N$ indicates the dimensionalities associated with these respective data categories. 
Additionally, within the \llmmoename{} framework, upon receiving supplementary text-based descriptions of user occupation and status, the \llmname{} processes this information to identify the most suitable expert model for the subsequent health alert decision-making process. 

\subsubsection{Health Alert Decision-Making}
In our \systemname{} framework, to train a neural network in the health decision-making task using the \drlname{}, we define the {\em state} as ${\boldsymbol s}=\{{\boldsymbol g}, {\boldsymbol d}, {\boldsymbol a}\}$, which includes the human postural profiles and the collected data. 
${\boldsymbol a}=[a_1, a_2, \ldots, a_K]^T$ represents the current human postural profiles, as detected by the \ddpgname{} algorithm using the mixed-content data pre-processed above, where $K$ denotes the number of users.
The {\em action}, represented as a vector, determines each physiological parameter's contribute weight, where a zero value negates the respective physiological parameter's influence. 
The {\em reward} mechanism within the \alertname{} system is strategically designed based on the Key Performance Indicators (\kpiname{}) to ensure that the health alert decision results align with the \qoename{} expectations of users. 
In this study, the reward is quantitatively defined as the negative of the combined probabilities of \maname{} and \faname{}, represented mathematically as $-\{{\textrm P}(\mbox{MA})+{\textrm P}(\mbox{FA})\}$.

\begin{figure*}[!t]
\centering
\includegraphics[width=\textwidth]{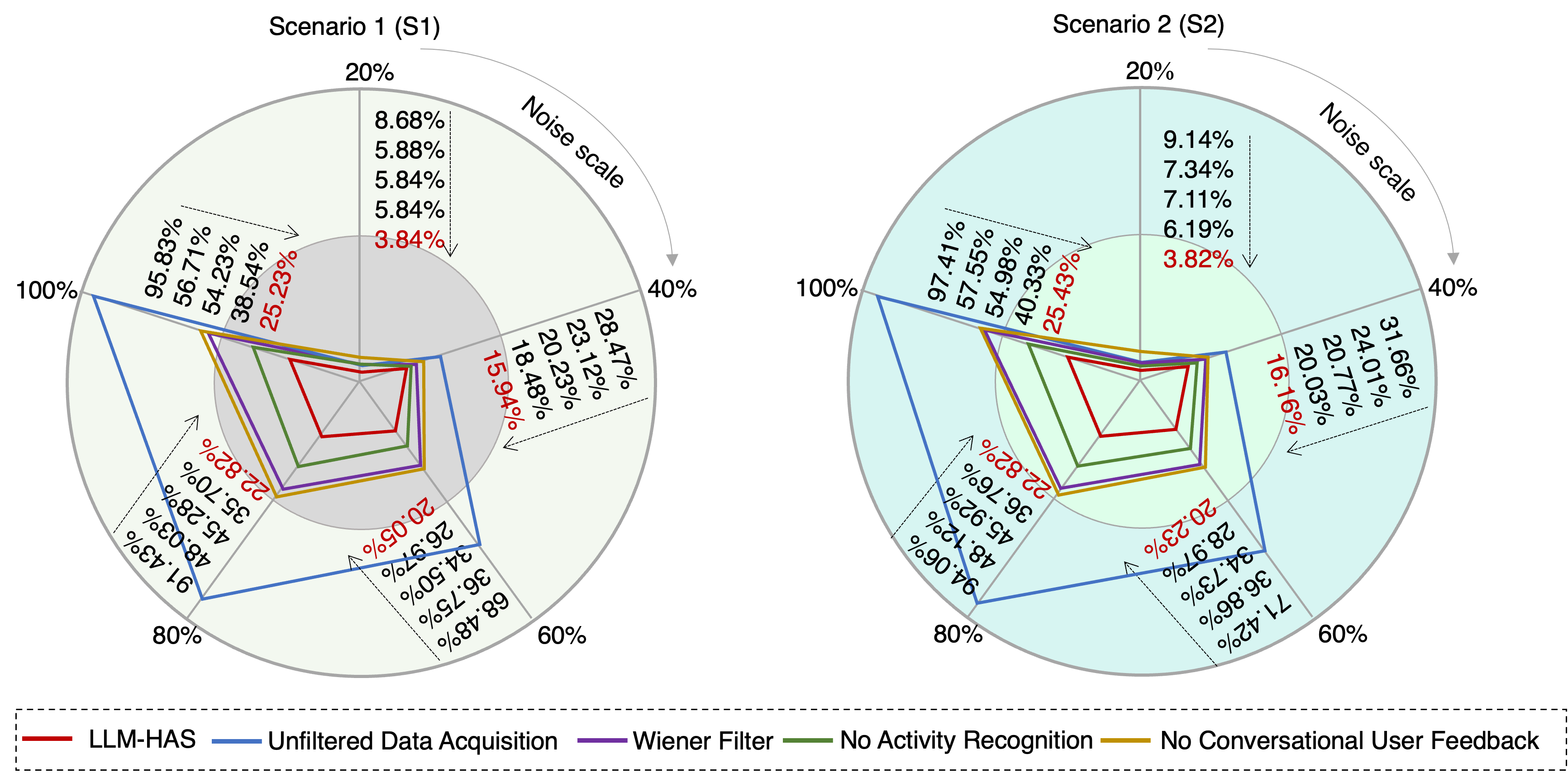}
\caption{Performance comparison of the probability of \faname{} and \maname{} versus different level of noise under scenario 1 and scenario 2.}
\label{fig:4}
\end{figure*} 
\subsubsection{Conversational User Feedback of \systemname{}}
Given the context of a novel \alertname{} framework that utilizes \llmname{} for natural language processing on the user side, the proposed conversational user feedback can accurately reflect the performance and user subjective QoE of the alert decisions. 
To clearly demonstrate the conversational user feedback mechanism within the \systemname{} framework, an example is presented in Fig. \ref{fig:3}. 
The process initiates with a smart bracelet collecting heart rate and breathing rate data. 
If these metrics are unusually high, the system automatically issues warnings. 
In this scenario, the user, engaged in intense physical activity like running, provides voice feedback through the device: {\bf\em \{I am running, and my heart rate and breath rate are normal for this activity; I don’t feel any discomfort.\}}
This conversational feedback prompts the edge server to finetune the \ddpgname{} model, possibly by increasing the threshold for triggering health alerts. 
The model reassesses the situation and determines whether the initial warnings remain relevant. 
Based on this refined analysis, it might retract previous warnings or offer additional suggestions for the user. 
The user then reviews and positively acknowledges the system’s adjusted decision-making, enhancing the model's accuracy and relevance for future interactions.
For a user to give a comprehensive and subjective evaluation of an alert decision in a \alertname{}, several aspects should be considered to ensure that the feedback is thorough and actionable. 
Here are key aspects that users should evaluate and provide feedback on:

\begin{itemize}
\item{\bf Accuracy of the alert.} 
{i) Relevance: Ensuring the alert is relevant to the user's current health situation or medical history is crucial for the system's effectiveness. Irrelevant alerts could lead to alarm fatigue or distrust in the system.}
{ii) Precision: The alert must accurately reflect the severity and nature of the health issue. Accurate alerts help in making appropriate health decisions, enhancing user safety.}
\item{\bf Timeliness. }
{ Timely delivery of alerts is critical in healthcare settings, where delayed information can result in missed opportunities for early intervention. Ensuring that alerts are prompt can significantly impact the management of potential health issues.}
\item{\bf  Clarity and understandability.}
{The information provided in the alert should be clear and easy to understand. This ensures that users can comprehend what the alert is about and what actions, if any, need to be taken. Clear communications is vital to avoid confusion and ensure effective response. }
\item{\bf Impact on user.}
{ Evaluating how the alert impacts the user--whether it provides valuable health management information without causing unnecessary anxiety--is essential. 
The goal is to support the user's health management effectively without adding stress or discomfort.}
\item{\bf Suggestions and improvement.}
{This feedback is crucial for ongoing refinement of the system. Understanding what additional features users desire or what adjustments could enhance their experience helps tailor the system to better meet user needs. }
\end{itemize}

Encouraging users to consider these aspects in their evaluations can provide healthcare providers, specifically those managing the edge server, with valuable insights into the effectiveness and user-friendliness of the \systemname{}.
In this article, feedback is facilitated through straightforward methods like direct textual and voice evaluations of various indicators, allowing users to express their subjective feelings easily. 
By implementing this \systemname{} framework, the system can be continuously refined and improved to more effectively meet user needs. 

\section{Experiments of \systemname{}}\label{sec:simulations}
In this section, we conduct extensive experiments to evaluate the \systemname{} based on the dataset from the {\bf{UC Irvine Machine Learning Repository}} \cite{misc_mhealth_319}. 

\subsection{Experimental Setting}
The MHEALTH dataset comprises body motion and vital signs recordings for ten volunteers of diverse profile while performing several physical activities.
Sensors placed on the subject's chest, right wrist and left ankle are used to measure the motion experiences by diverse body parts, i.e., acceleration, rate of turn and magnetic field orientation.
The sensor positioned on the chest also provides 2-lead ECG measurements, which can be potentially used for basic heart monitoring, checking for various arrhythmias or looking at the effects of exercise on the ECG.
Additionally, the human postural dataset includes postures such as standing, sitting, lying, walking, climbing stairs, waist bends forward, frontal elevation of arms, knees bending, cycling, jogging, running, and jumping.
Finally, the edge server is equipped with an Intel i7-13700KF processor, 32 GB of 6600 MHz DDR5 RAM, and an Nvidia RTX-4090 graphics card.

To evaluate the \systemname{} framework, particularly its diffusion-based denoising process and conversational user feedback mechanism, we select four relevant baselines for assessment. 
\begin{itemize}
\item[-] {\bf Unfiltered Data Acquisition}: The data collected by the edge server is not preprocessed.   
\item[-] {\bf No Activity Recognition}: the \systemname{} does not consider the impact of user activity on health alert decision. 
\item[-] {\bf No Conversational User Feedback}: the system operates without integrating interactive feedback from users, the health alert strategy is a one-time outcome.  
\item[-] {\bf Wiener Filter\cite{matthews2023advances}}: Wiener Filter is employed as a signal processing method to improve the accuracy and reliability of the collected data in \alertname{}.
\end{itemize}

To gain insights into the proposed \systemname{} framework, we conducted performance comparisons within two experimental setups: {\bf Scenario 1 (S1):} the noise is uniformly distributed on the interval $[-\delta\bar{\bigstar}/2, \delta\bar{\bigstar}/2]$, where $\bar{\bigstar}$ denotes the mean value of the original data $\bigstar$ collected by edge server, and $\delta$ represents the noise scale.  
{\bf Scenario 2 (S2):} the noise follows a Gaussian distribution, denoted as ${\mathcal C}{\mathcal N}(0,\delta^2)$.  
\begin{figure}[!t]
\centering
\includegraphics[width=3.8in]{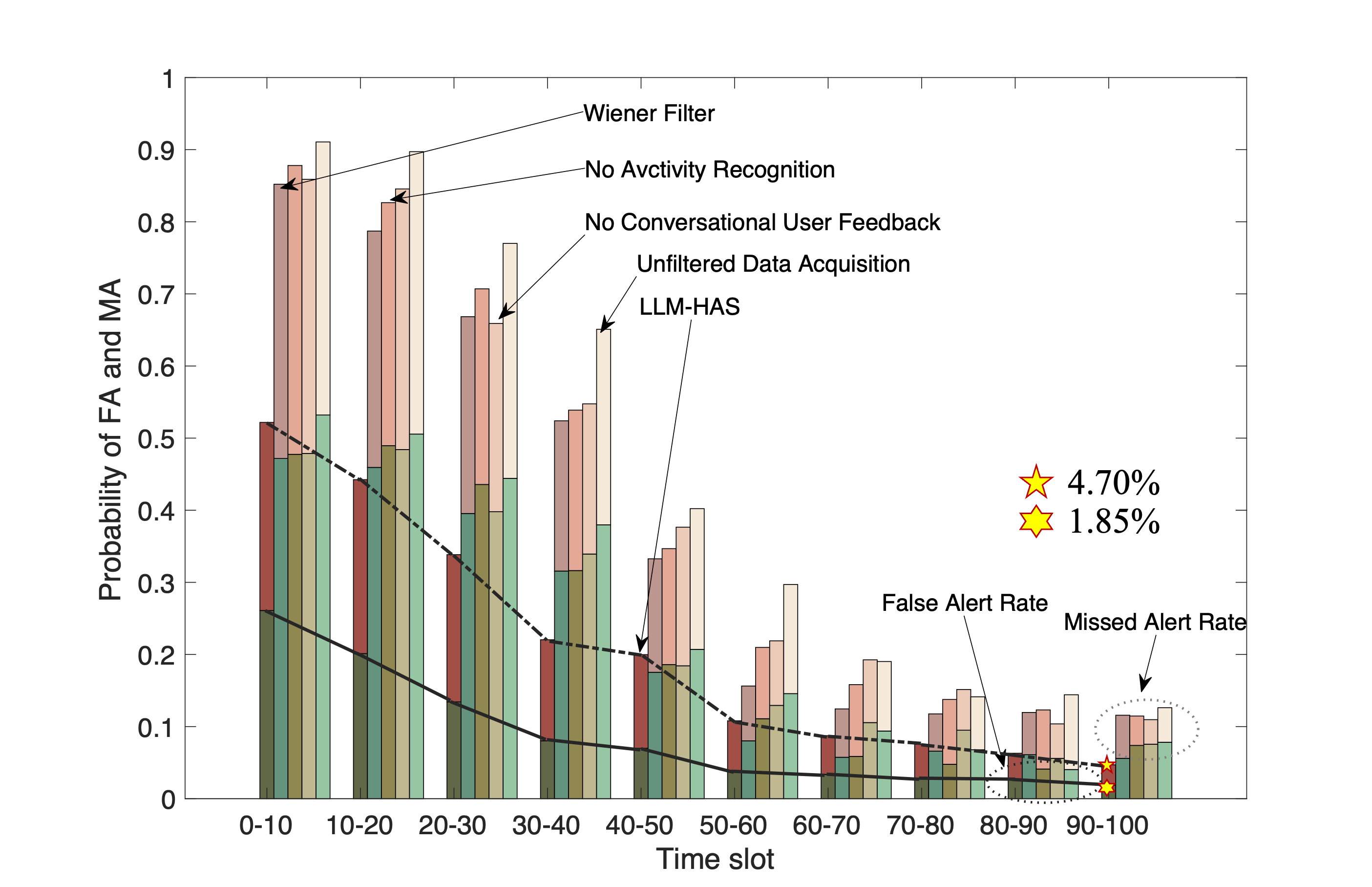}
\caption{Comparison of performance in terms of FA rate and MA rate, respectively. }
\label{fig:5}
\end{figure}

\subsection{Results and Discussion}
To begin with, we evaluate the advancement of our proposed diffusion-based denoising process by contrasting it with the Wiener Filter. 
As illustrated in Fig. \ref{fig:2}, it is difficult to distinguish between the curves of the pre-processed data by the diffusion model and the original data. 
In contrast, there is a significant gap between the curves of the Wiener Filter and the original data. 

After validating the diffusion-based denoising process, we continue to assess the performance of the comparison methods, specifically focusing on the probability of \faname{} and \maname{} in S1 and S2.
Fig. \ref{fig:4} illustrates the impact of varying noise levels on the probability of \faname{} and \maname{} under both S1 and S2, with $\delta$ ranging from 20\% to 100\%.  
From the results, we observe that the probability of \faname{} and \maname{} for all comparison methods rises with the increasing noise level $\delta$ in both S1 and S2.
Our \systemname{} consistently outperforms the alternatives and demonstrates exceptional robustness. 
As expected, the Unfiltered Data Acquisition method exhibits the poorest performance in health alert decision-making at any noise level in both scenarios. 
Furthermore, in both S1 and S2, the No Conversational User Feedback method ranks as the second poorest across all noise levels. 
In the Wiener Filter method, the collected data is preprocessed and conversational user feedback is incorporated, resulting in a lower probability of \faname{} and \maname{} compared to the Unfiltered Data Acquisition method and the No Conversational User Feedback method. 
In contrast, while the No Activity Recognition method performs best among the baselines, our \systemname{} reduces the probability of \faname{} and \maname{} by an average of 8.147\%. 

In order to fully demonstrate and prove the effectiveness of our \systemname{} framework, Fig. \ref{fig:5} shows the performance of both the \faname{} rate and \maname{} rate, and compare the changes of them with different methods.
From the results, we observe that both the \faname{} rate and the \maname{} rate for all methods decrease and then stabilize as time progresses. 
More precisely, for time slots of 40 or fewer, both the \faname{} rate and \maname{} rate significantly decrease as time progresses. 
Subsequently, in our \systemname{}, the \faname{} rate and \maname{} rate gradually stabilize at 1.85\% and 2.85\%, respectively, resulting in a total probability of \faname{} and \maname{} at 4.70\%. 

\section{Future Directions}\label{sec:future}
\subsection{Autonomous \systemname{} Framework}
Our initial \systemname{} framework enhances the accuracy of health alert decisions by fine-tunning the algorithm based on conversational user feedback. 
Building on this foundation, the primary goal of developing an autonomous  \llmname{}-enabled smart healthcare framework is to evolve these capabilities into a system that can operate independently.
Specifically, we plan to leverage the advanced capabilities of \llmsname{} to more efficiently understand users' intentions and generate respective code. 
This advanced system would not only retain its ability to make informed decisions about health alerts but also adapt to new data.
These enhancements will ensure that LLM-based system not only delivers timely health interventions but also aligns with the evolving needs of smart healthcare infrastructure. 

\subsection{Energy-Efficient Wireless Transmission Methods}
Investigate and implement energy-efficient transmission methods that reduce the power consumption of IoT devices, thereby extending their battery life and ensuring continuous operation.
This includes strategies like adaptive transmission power control, energy-harvesting technologies, and optimized network protocols for ``asymmetric processing" between the edge server and IoT devices, where the former processes most of the requirements for energy-efficient management of IoT devices.
Simultaneously, we may consider the ``semantic communication (SemCom)" for delivering the minimal necessary information relevant to a specific task associated with a target IoT device. 
Especially, the SemCom helps to boost the protection of user data privacy as studied in \cite{du2023rethinking}. 
These improvements not only boost the sustainability and efficiency of health monitoring systems but also enhance their reliability, ensuring that healthcare providers receive timely and accurate data for informed decision-making, thus leading to better healthcare outcomes.

\subsection{Human Behavior Consistency}
Behavioral consistency in \llmname{} ensures that the model's actions align with its intended task goals. 
While current research has adapted \llmsname{} to mirror typical human behaviors, applying these models in the medical field is more complex. 
Medical professionals operate under strict clinical guidelines and ethical standards that differ greatly from general human behavior. 
Therefore, implementing \llmsname{} in healthcare requires specific adjustments. 
This includes creating specialized training data, enhancing the model's understanding of medical contexts, and integrating it with clinical support systems to ensure accuracy and professionalism. 
Such steps are essential for \llmsname{} to effectively contribute to improved diagnostics, patient management, and healthcare services.

\section{Conclusions}\label{sec:conclusion}
This article proposed the \systemname{} framework by incorporating the \llmname{} and \gdmname{} into the typical \alertname{}. 
The incorporation of \llmsname{} allowed for the dynamic reconstruction of sensitive data, which was then securely processed and analyzed using a GDM and the \ddpgname{} algorithms to improve decision-making in health alerts. 
The inclusion of \llmname{}-enabled Conversational User Feedback enriches user interaction, significantly enhancing the subjective QoE and ensuring that user input directly influences system outputs. 
The active user involvement not only refined the system's performance but also boosted trust and user satisfaction by tailoring alerts to individual demands and contexts. 
Finally, numerical simulation results confirmed the efficacy of the \systemname{} framework, showcasing substantial improvements in the accuracy and reliability of health alerts. 
To the best of our knowledge, this is the first \systemname{} framework to incorporate a diffusion-enabled denoising process and \llmname{}-enabled Conversational User Feedback specifically designed for health alert systems. 

\bibliography{reference}
\bibliographystyle{IEEEtran}

\end{document}